# Deep learning-enabled virtual multiplexed immunostaining of label-free tissue for vascular invasion assessment

Yijie Zhang[†,1,2,3], Çağatay Işıl[†,1,2,3], Xilin Yang[1,2,3], Yuzhu Li[1,2,3], Anna Elia[4], Karin Atlan[4], William Dean Wallace[5], Nir Pillar[*,1,2,3,4], and Aydogan Ozcan[*,1,2,3,6]

[1]Electrical and Computer Engineering Department, University of California, Los Angeles, CA, 90095, USA.

[2]Bioengineering Department, University of California, Los Angeles, CA, 90095, USA.

[3]California NanoSystems Institute (CNSI), University of California, Los Angeles, CA, 90095, USA.

[4]Department of Pathology, Hadassah Hebrew University Medical Center, Jerusalem, 91120, Israel

[5]Department of Pathology, Keck School of Medicine, University of Southern California, Los Angeles, CA, 90033, USA.

[6]Department of Surgery, University of California, Los Angeles, CA, 90095, USA.

*Correspondence: Nir Pillar nir.pillar@mail.huji.ac.il, Aydogan Ozcan, ozcan@ucla.edu

[†] Equal contributing authors

## Abstract

Immunohistochemistry (IHC) has transformed clinical pathology by enabling the visualization of specific proteins within tissue sections. However, traditional IHC requires one tissue section per stain, exhibits section-to-section variability, and incurs high costs and laborious staining procedures. While multiplexed IHC (mIHC) techniques enable simultaneous staining with multiple antibodies on a single slide, they are more tedious to perform and are currently unavailable in routine pathology laboratories. Here, we present a deep learning-based virtual multiplexed immunostaining framework to simultaneously generate ERG and PanCK, in addition to H&E virtual staining, enabling accurate localization and interpretation of vascular invasion in thyroid cancers. This virtual mIHC technique is based on the autofluorescence microscopy images of label-free tissue sections, and its output images closely match the histochemical staining counterparts (ERG, PanCK and H&E) of the same tissue sections. Blind evaluation by board-certified pathologists demonstrated that virtual mIHC staining achieved high concordance with the histochemical staining results, accurately highlighting epithelial cells and endothelial cells. Virtual mIHC conducted on the same tissue section also allowed the identification and localization of small vessel invasion. This multiplexed virtual IHC approach can significantly improve diagnostic accuracy and efficiency in the histopathological evaluation of vascular invasion, potentially eliminating the need for traditional staining protocols and mitigating issues related to tissue loss and heterogeneity.

# Introduction

The introduction of immunohistochemistry (IHC) into routine pathology practice in the early 1990s marked a significant advancement in diagnostic precision. By enabling the visualization of specific proteins within tissue sections, IHC allowed pathologists to move beyond morphological assessment and incorporate molecular information into their diagnoses[1]. In routine clinical pathology, IHC typically requires one slide per antibody stain, as each marker is applied to a separate tissue section. This one-marker-per-slide approach results in increased tissue consumption, especially critical in small biopsies. It also introduces variability due to section-to-section differences and increases the workload and costs associated with slide preparation, staining, and interpretation.

While standard IHC methodology enables labeling of a single marker per tissue section, multiplexed IHC (mIHC) technologies, introduced over the last decade, allow simultaneous staining of a tissue section with multiple antibodies[2]. These multiplexed histochemical staining approaches, however, demand advanced imaging and analysis tools, are expensive and have a very slow turnaround time. None of these methodologies has received clinical approval as diagnostic assays. To overcome these limitations while maintaining compatibility with existing clinical workflows, pathology labs have adopted duplex IHC staining—a method that uses two antibodies, each detected by a distinct chromogen through secondary antibodies. This approach enables spatial localization of two biomarkers on the same tissue slide using standard IHC protocols routinely employed in clinical practice. One such example is the identification of vascular invasion, defined by the presence of tumor cells within blood vessels or lymphatic vessels. Vascular invasion represents a critical step in the metastatic cascade, facilitating both lymphatic spread and systemic dissemination of cancer cells. Its prognostic significance has been well established across a range of solid tumors, including breast[3], colon[4], bladder[5], lung[6], thyroid carcinoma[7] and melanoma[8]. The identification of vascular invasion may influence clinical decision-making and support the implementation of more aggressive approaches, such as adjuvant chemotherapy or radiation.

The identification of vascular invasion on hematoxylin and eosin (H&E) stained sections is an important component of histopathologic evaluation of tumors, but it can be challenging due to artifacts such as tissue retraction, cracks, or pseudovascular spaces[4] that may mimic true vascular structures. While IHC staining for endothelial markers such as ERG, CD31, or D2-40 can help, they do not provide information about the nature of the intravascular cells. In such cases, uncertainty may arise as to whether the cells within the vessel are malignant or represent benign cells, such as histiocytes. To address this limitation, duplex IHC staining, which combines an endothelial marker (e.g., ERG) with an epithelial marker (e.g., cytokeratin), can allow for the simultaneous visualization of both vascular architecture and tumor cell localization, thereby improving diagnostic accuracy and reducing the risk of misinterpretation[9,10]. However, the duplex IHC technique is available only in large pathology labs, whereas most small- to medium-sized laboratories typically perform a single IHC stain per slide, as illustrated in Figure 1(a). In certain cases where vascular invasion is suspected on H&E-stained sections and IHC is ordered to confirm the presence of tumor cells within vascular structures, the area of interest may be lost in deeper tissue levels (particularly when involving small-caliber vessels) due to sectioning variability or tissue dropout. This challenge underscores the importance of performing immunostaining on sections cut at the same tissue level as the original H&E slide to ensure accurate localization and interpretation of the suspicious focus.

Recently, virtual histological staining of label-free tissue sections has been successfully demonstrated for various stain types[11–23]. These techniques computationally transform microscopy images of unstained tissue into digitally stained images that closely match conventional chemically stained images of the same tissue samples. Advances in deep learning methods and powerful computing hardware have significantly accelerated the development of these virtual staining approaches[16,18–21,24–33], effectively applying them to common stains such as H&E[11] and IHC[34]. Recent advancements in multiplexed virtual staining have enabled the visualization of multiple stains (e.g., H&E, Masson's trichrome, Jones' silver) on the same tissue area using a single trained model[35]. This multiplexing approach was further extended to concurrently learn cross-modality image transformations within a single neural network, simultaneously generating virtual birefringence imaging and virtual Congo red staining of amyloid deposits from the same label-free autofluorescence inputs[36].

Here, we present a deep learning-based virtual multiplexed immunostaining approach designed to simultaneously generate virtual H&E and duplex IHC staining on label-free thyroid tissue sections, enabling accurate localization and interpretation of vascular invasion in thyroid cancers. As illustrated in Figure 1(b), our method utilizes conditional generative adversarial networks (cGANs) to rapidly convert autofluorescence (AF) microscopy images of unstained tissue slides into virtually stained images (H&E, ERG, and PanCK), matching the corresponding histochemically stained images of the same tissue section. To accomplish multiplexed staining, we integrated a digital staining matrix as an additional input channel alongside autofluorescence images within the cGAN framework, as depicted in Figure 1(c). To validate this approach, we trained our virtual staining network using a dataset mixed with paired autofluorescence images of label-free tissue microarray (TMA) cores and their histochemically stained counterparts (H&E, ERG, and PanCK). After its training, our virtual mIHC staining network successfully generated virtual H&E, ERG, and PanCK-stained images from label-free test TMA cores, never used during training. The resulting virtual IHC images (ERG and PanCK) and their corresponding histochemically stained ground truth were randomly shuffled and blindly evaluated by board-certified pathologists. Four pathologists independently affirmed the non-inferior quality of the digitally generated virtual duplex IHC images compared to their histochemical staining counterparts, noting a high concordance with the ground truth images. Additionally, a pathologist successfully identified and localized vascular invasion using the generated virtual H&E and duplex IHC images. The presented virtual mIHC staining approach effectively eliminates the need for H&E staining followed by dual IHC staining procedures on serial tissue sections, circumventing limitations such as tissue dropout and sectioning variability, and it has the potential to accelerate and enhance the diagnostic accuracy for vascular invasion in routine histopathological evaluation.

# Results

**Virtual mIHC staining of label-free tissue**

Initially, we captured label-free autofluorescence microscopy images from TMA cores of human thyroid tissue samples. Subsequently, these samples were separately subjected to conventional histopathological staining processes, including H&E and IHC staining with ERG and PanCK antibodies. The resulting stained sections were scanned using brightfield microscopy to obtain brightfield (BF) images of stained TMA cores, serving as ground-truth references for both the training and testing phases. Each AF image of label-free TMA core was paired and finely registered with its corresponding BF image after each staining process (one staining modality per section), resulting in three distinct datasets corresponding to the H&E, ERG, and PanCK stains; see the Methods section for details. Our virtual mIHC staining model leveraged a conditional generative adversarial network, incorporating a digital staining matrix (DSM) concatenated with the input autofluorescence channels. This DSM, matching the pixel dimensions of the input images, determined the desired staining output at the pixel level, encoded as "2" for H&E, "1" for PanCK, and "-1" for ERG staining. During the training process, all three datasets were mixed to ensure the network learned accurate transformations from label-free AF images to the respective virtual staining channel. Further details on data acquisition, preprocessing, and network architecture are presented in the Methods section.

Following its training, the virtual mIHC staining model was evaluated in a blinded manner using label-free AF images from 30 previously unseen TMA cores specifically reserved for testing. Among these 30 cores, 12 were subsequently stained with H&E, 12 with PanCK, and 6 with ERG to acquire ground-truth histochemically stained images – only used for comparison purposes. Across all three staining modalities, the generated virtual staining images exhibited a good agreement with the corresponding ground truth images, as illustrated in Figure 2. This figure provides side-by-side visual comparisons between the virtually stained images produced by our model and their histochemically stained counterparts. Specifically, Figures 2(a, c, e) demonstrate representative examples from three distinct patient-based TMA cores (not used for model training), each subjected to H&E, PanCK, or ERG staining after the autofluorescence imaging. Detailed comparisons of the zoomed-in regions for each staining modality are presented in Figures 2(b, d, f), highlighting the quality of the virtual staining results relative to the histochemically stained ground truth images. The virtual H&E slides effectively replicate the tissue morphology and color contrast observed in the histochemical reference, allowing for the evaluation of cell size and shape, nuclear chromatin texture, and stromal organization. Virtual IHC staining for PanCK shows cytoplasmic labeling of epithelial cells, enabling epithelial distribution and structural integrity. As highlighted by the red-circled regions in Figure 2(f), the virtual ERG IHC staining exhibits nuclear patterns, permitting the evaluation of vascular and endothelial components, demonstrating high concordance with the histochemical ground truth images.

## Evaluations of virtual mIHC (PanCK and ERG) staining by board-certified pathologists

To further validate the efficacy of our deep learning-based mIHC virtual staining approach, we conducted a study involving 39 pairs of virtual ERG images with their corresponding histochemically stained ground truth images, as well as 46 pairs of virtual PanCK images with their corresponding ground truth images. The image pairs (each image with 2000×2000 pixels) were sampled from 6 ERG-stained TMA cores and 6 PanCK-stained TMA cores reserved for testing. As illustrated in Figure 3, these 85 image pairs (totaling 170 images) were randomized and sequentially presented in a blinded manner to three board-certified pathologists (A.E., K.A., and N.P.). Each pathologist evaluated the images based on four standardized questions:

1. What is the staining pattern? ("n" for nuclear or "c" for cytoplasmic)
2. How would you rate the staining intensity? (Scale: 1 for weak, 2 for moderate, and 3 for strong)
3. Indicate your agreement with the statement:
   - For nuclear staining patterns: "*The stain highlights blood vessels.*"
   - For cytoplasmic staining patterns: "*The stain highlights epithelial cells.*"
4. Indicate your agreement with the statement:

- o For nuclear staining patterns: "*The stain highlights cells other than endothelial cells*."
- o For cytoplasmic staining patterns: "*The stain highlights cells other than epithelial cells*."

For questions 3 and 4, responses were recorded on a three-point scale: 1 for "disagree," 2 for "equivocal," and 3 for "agree." Additionally, if a pathologist was unable to confidently respond to any question, their answer was recorded as "cannot determine." After obtaining evaluations from the three pathologists, we applied a majority voting strategy to determine the final response for each image, as depicted in Figure 3. If all three responses differed (resulting in a tie), an additional evaluation was performed by a senior $4^{th}$ pathologist (W.D.W) to provide a final adjudicated response.

The evaluation results for the 46 pairs of virtual and histochemical PanCK-stained images, assessed according to predefined evaluation questions, are summarized in Figure 4(a–d). These results illustrate strong concordance between virtually stained PanCK images and their histochemically stained counterparts across multiple evaluation dimensions. Specifically, the pathologists consistently recognized both virtual and histochemically stained PanCK images as having a robust cytoplasmic staining pattern, effectively highlighting the targeted epithelial cells without significant staining of off-target cells. To better compare the virtual PanCK images with their histochemical counterparts, Figure 4(e) presents a statistical superiority analysis based on the pathologists' responses to the four evaluation questions. The virtually stained images consistently matched the histochemical staining results, and notably, in some instances, virtual staining demonstrated superior performance, particularly regarding consistent stain intensity and targeted epithelial cell highlighting.

Similarly, the evaluation results for the 39 pairs of ERG-stained images are presented in Figure 5(a–d). The majority of pathologists recognized both the virtual and histochemical ERG-stained images as having a nuclear staining pattern, consistently highlighting targeted endothelial cells associated with blood vessels. Figure 5(e) provides a statistical superiority analysis, again indicating a high level of agreement between the virtual and histochemically stained tissue images across all evaluation aspects. Notably, the virtual ERG staining occasionally exhibited superior performance, particularly regarding staining specificity with minimal background staining. This advantage potentially arises from technical limitations inherent in histochemical ERG staining, such as non-specific antibody binding and suboptimal antibody quality, which can sometimes result in inadequate visualization of endothelial cells. Thus, the virtual mIHC staining method demonstrates superior reliability in consistently labeling endothelial cells.

To further assess the relevant histopathological features highlighted in this comparison, a board-certified pathologist (N.P.) reviewed virtual and histochemical ERG images for three representative cases (#7, #19, and #21). In cases #7 and #21, where the virtual ERG staining was superior, the virtually stained sections clearly highlighted several endothelial cells with strong nuclear staining, while the corresponding histochemical stains failed to detect ERG-positive endothelial cells. Conversely, in case #19, the histochemical staining was superior as multiple endothelial cells displayed positive ERG staining, whereas the virtual staining highlighted only occasional ERG-positive cells.

These pathological evaluation results, both quantitatively and qualitatively, demonstrate that the virtually stained PanCK and ERG images effectively highlight target epithelial cells and endothelial cells, respectively. To further illustrate how our virtual mIHC staining assists in identifying vascular invasion, Figure 6(a–c) presents virtually stained H&E, PanCK, and ERG images of a representative TMA core from a patient diagnosed with papillary thyroid carcinoma and distant metastasis. The corresponding histochemically stained PanCK ground-truth image is also displayed in Figure 6(d). The yellow arrows in the zoomed-in images highlight endothelial cells (Figure 6(f)) and epithelial cells (Figure 6(g)) within the vessel lumen, indicative of vascular invasion, as identified by a board-certified pathologist (N.P.). In contrast, the histochemically stained PanCK image exhibited some staining artifacts, as illustrated by the blue-circled region in Figure 6(h), thereby failing to provide accurate target identification. These results underscore the potential impact and effectiveness of our virtual staining technique in reliably localizing and identifying vascular invasion.

# Discussion

Accurate identification of vascular invasion, defined as the presence of tumor cells within blood or lymphatic vessels, is pivotal for determining cancer prognosis and guiding clinical management due to its critical role in facilitating metastatic dissemination. Conventional histopathological assessment typically starts with H&E-stained sections, aimed at concurrently visualizing the entire tissue composition, including tumor cells and vascular structures. However, identifying blood vessels, and specifically, small-caliber vessels, can be confounded by artifacts such as tissue retraction, cracks, or pseudovascular spaces, complicating the reliable distinction between true vascular structures and artifact-induced formations. This leads to a significant number of cases where the vascular lumen cannot be definitely identified in the H&E section[37,38]. To mitigate these limitations, routine clinical workflows often incorporate additional IHC stains such as ERG to confirm endothelial lining presence and PanCK to validate tumor cell localization within the suspected lumen space, typically performed on serial tissue sections, as illustrated in Figure 1(a). However, this sequential staining process poses challenges, notably the potential loss of critical areas of interest—especially those involving small-caliber vessels—due to tissue dropout or because the examined H&E-stained section represents the final level at which the vessel is still present. Multiplexed staining methods, which enable simultaneous detection of multiple biomarkers, have emerged to directly address multiple antigen localization on the same tissue slide. Nonetheless, none of these advanced methods are incorporated into routine diagnostic workflows as they typically require specialized imaging equipment, complex analytical frameworks, substantial financial resources, and prolonged tissue processing times. Furthermore, multiplexed staining methods are typically performed on sections sequential to the H&E-stained slide. If a vessel is present only on the H&E section and absent at deeper levels, it will not be detected using these methods.

In this study, we introduced a virtual mIHC framework designed to transform label-free autofluorescence microscopy images of thyroid tissue sections into brightfield equivalents of H&E, ERG, and PanCK stains. This framework demonstrated strong concordance with traditional histochemical staining results. Importantly, simultaneous generation of multiple stains (H&E and two different IHCs) on a single physical tissue section, as achieved here, is unattainable through current standard histopathological methods, and this capability allows for the definite detection of small-caliber vessels present only on the initial section. Our approach ensures the accurate co-localization of tumor cells and blood vessels within the same physical tissue section, significantly enhancing diagnostic accuracy by eliminating issues such as tissue dropout or the possibility that the examined H&E-stained section represents the last level where the vessel is still visible, thus preserving areas of interest and improving diagnostic efficiency.

Our virtual mIHC method requires only a single tissue section subjected to autofluorescence imaging, circumventing the cumbersome and labor-intensive process of handling serial tissue sections and subsequent individual histochemical staining steps. This can streamline clinical workflows and conserve valuable tissue for subsequent advanced molecular analyses. Autofluorescence microscopy was chosen as the imaging input modality due to its proven effectiveness in multiple virtual staining studies[11,26,27,34,36] and its ease of integration with existing FDA-cleared whole-slide imaging systems[39,40] without additional optical components or significant costs. After a one-time training phase—including data acquisition, preprocessing, and network optimization—our multiplexed virtual staining framework can rapidly generate virtual mIHC images. Specifically, it completes inference in mere seconds for individual fields (e.g., <2 seconds for a 2000×2000 pixel field-of-view) and in a few minutes for a whole-slide image when leveraging modest graphics processing units (GPUs).

The robustness of our method was validated through a blinded evaluation by four board-certified pathologists, confirming that virtually stained ERG and PanCK images accurately highlighted targeted endothelial and epithelial cells, respectively. In the study, we did not focus on H&E image quality comparisons because virtual H&E staining using label-free microscopy (e.g., autofluorescence microscopy) and cGANs has been extensively demonstrated to be robust and physiologically equivalent to the histochemical H&E counterparts in previous efforts[11,12,14,15,17,20–22,26,41,42]. Notably, the statistical superiority analysis demonstrated in Figures 4(e) and 5(e) revealed that virtual staining results in general provided superior consistency in staining intensity and specificity. This comparative analysis highlights another significant advantage of our virtual mIHC staining model—its stability and reproducibility—delivering consistently high-quality staining images. Conversely, conventional histochemical staining frequently encounters various technical artifacts, such as uneven staining quality, overstaining, understaining and pigment deposition leading to morphological alterations and diagnostic challenges, as illustrated in Figure 6(h). Furthermore, traditional IHC staining sometimes introduces artifacts like non-specific antibody binding and variability in antibody quality, contributing to inadequate or inconsistent visualization of targeted cells, exemplified by the weaker staining intensity observed in histochemical ERG images (Figure 5).

Our results, depicted in Figure 6, further demonstrate that the virtual mIHC approach effectively identifies and localizes vascular invasion. Within the same tissue section, pathologists can initially perform routine cancer assessments using the virtual H&E image to identify suspicious regions. Virtual ERG and PanCK images, at the same nanoscopic grid of the same tissue section, can subsequently highlight endothelial cells around vessels and epithelial cells, respectively, enabling accurate identification of vascular invasion. In contrast, traditional diagnostic workflows risk losing the region of interest due to tissue dropout or the vessel not being present on deeper levels, even after initially pinpointing suspicious areas on H&E-stained sections. Our virtual mIHC staining framework provides a reliable and consistent solution, potentially improving diagnostic accuracy and workflow efficiency.

The integration of a digital staining matrix with the virtual staining model ensures internal consistency across the generated staining types, maintaining structural alignment between the virtual mIHC and H&E outputs. Conversely, employing separate neural networks for each virtual stain could introduce divergent artifacts and network-to-network inconsistencies, potentially complicating clinical interpretation. An exciting future research direction could involve automated label-free detection of potential vascular invasion areas, subsequently applying multiplexed *microstructured* staining, e.g., ERG around blood vessels, PanCK within vessels, and H&E in surrounding regions. Such multiplexed microstructured staining could further assist pathologists by accelerating diagnostic workflows.

In conclusion, our virtual mIHC method effectively transforms label-free thyroid tissue sections into brightfield microscopy images equivalent to histochemical H&E, ERG, and PanCK stains. This method facilitates rapid and accurate localization and identification of vascular invasion, eliminating the need for IHC staining procedures on serial tissue sections and reducing the risk of losing areas of interest due to tissue dropout or heterogeneity. Our approach holds promise for significantly transforming conventional diagnostic workflows used for vascular invasion, paving the way for extensive, multi-center clinical validation studies to further advance the clinical translation and adoption of the presented technology.

## Methods

### Sample preparation, image acquisition, and histochemical staining

10 unlabeled, anonymized thyroid tissue microarray slides, each containing ~80 cores, were acquired from TissueArray[43]. Following autofluorescence imaging, two slides were sent for standard H&E staining, three for IHC PanCK staining, and four for ERG staining. All histochemical staining procedures were performed at the University of Southern California (USC) Pathology Lab. The autofluorescence images of label-free thyroid core biopsies were captured using a Leica DMI8 microscope with a 40×/0.95 NA objective lens (Leica HC PL APO 40×/0.95 DRY), controlled using Leica LAS X microscopy automation software. Four fluorescence filter cubes, including DAPI (Semrock OSFI3-DAPI5060C, EX377/50 nm EM 447/60 nm), TxRed (Semrock OSFI3-TXRED-4040C, EX 562/40 nm EM 624/40 nm), FITC (Semrock FITC-2024B-OFX, EX 485/20 nm, EM 522/24 nm), and Cy5 (Semrock CY5-4040C-OFX, EX 628/40 nm, EM 692/40 nm), were used to acquire the four autofluorescence image channels. This image acquisition was completed using a scientific complementary metal-oxide-semiconductor (sCMOS) image sensor (Leica DFC 9000 GTC) with exposure times of 150 ms, 300 ms, 300 ms, and 500 ms for the DAPI, FITC, TxRed, and Cy5 filters, respectively. After the histochemical staining process, the tissue samples were imaged using a pathology scanner (Aperio AT2, Leica Biosystems, 20×/0.75NA objective with a 2× adapter) for capturing the ground truth BF images.

**Image preprocessing and registration**

We used an image registration strategy to precisely align the autofluorescence images and brightfield (ground truth) images. This is a critical step to train virtual staining models since we incorporated a supervised learning strategy for training these models. First, we stitched all the scanned images into the whole-slide images, including tissue microarray core biopsies for autofluorescence and brightfield imaging modalities. Subsequently, image pairs of each core were globally registered using a multi-modal registration algorithm[44,45]. Second, these roughly aligned autofluorescence and brightfield images for core biopsies were divided into small image tiles with 2000×2000 pixels. Finally, our data processing includes a correlation-based elastic registration step, which is critical to further match the image pairs with the local and global corrections and address optical aberrations of different imaging modalities and morphological distortions resulting from histochemical staining procedures. During this step, a registration neural network was first trained to translate the autofluorescence images into the brightfield stained versions (PanCK, ERG, and H&E) as well as elastically align brightfield images to the autofluorescence images during the training process[46]. This registration network module shared the same architecture and settings as our virtual staining network (see the "Neural network architecture" section). We further aligned the brightfield images into the style-transferred images (obtained using autofluorescence images) by using the pyramid elastic registration algorithm[11,47]. This algorithm includes pyramidal matching of local features across multi-resolution sub-image blocks to calculate transformation maps and then correction of the local distortions related to aberrations and staining procedures. This elastic registration step was repeated until very well-registered image pairs were obtained to enable accurate training of the virtual staining model. These registration algorithms were implemented using MATLAB (MathWorks) and Python.

**Dataset division and preparation**

After image processing and registration, we obtained three datasets consisting of paired autofluorescence images and their corresponding histochemically stained counterparts (PanCK, ERG, and H&E). For the H&E dataset, we used 4,416 image pairs (2000 × 2000 pixels for each tile) from 140 TMA cores for training and reserved 12 whole TMA cores (both autofluorescence and H&E-stained) for blind testing. For the PanCK dataset, we used 3,598 image pairs from 146 TMA cores for training and reserved 12 whole TMA cores for testing. For the ERG dataset, we used 1,216 image pairs from 70 TMA cores for training and reserved 6 whole TMA cores for testing. Due to technical limitations inherent to histochemical ERG staining—such as suboptimal antibody performance leading to poor visualization of endothelial cells—many ERG-stained cores with insufficient endothelial cell visibility were excluded from our ground truth images based on the evaluation of a board-certified pathologist (N.P.). As a result, the ERG staining dataset was approximately half the size of the H&E and PanCK datasets due to the relatively poor repeatability of the ERG histochemical staining. To address this class imbalance during training, we designed the data pipeline to ensure equal representation: in each batch, image pairs were randomly sampled from the three staining datasets with equal probability.

**Neural network architecture**

We used a cGAN architecture[48,49] to design a virtual mIHC staining model that rapidly transforms label-free autofluorescence images (DAPI, FITC, TxRed, and Cy5) of tissue samples into their corresponding brightfield images for ERG, PanCK, and H&E virtual staining. The cGAN architecture includes two subnetworks, namely a generator ($G$) and a discriminator ($D$). To enable the multiplexing feature for this virtual staining model, a digital staining matrix is concatenated with the label-free input images. As shown in Figure 7(a), the digital staining matrix $\tilde{c}$ is generated using one of three labels $\tilde{c} = [-1] \; or \; [1] \; or \; [2]$, which correspond to different virtual staining tasks, i.e., virtual ERG, PanCK, and H&E staining, respectively. For example, $\tilde{c} = [-1]$ denotes a digital staining matrix for ERG with all elements equal to -1. During the training phase, the generator was optimized to learn the virtual multiplexed staining process. In the meantime, the discriminator was trained to separate the generator outputs from ground truth stained images; see the Supplementary Information and Refs. 50-51 for the loss function-related details. The generator network (see Figure 7(b)), based on an Attention U-Net architecture[52], features a symmetrical encoder-decoder structure. The encoder path comprises four downsampling blocks, each containing a three-convolutional-layer residual block[53] (formed by three consecutive convolutional layers and a convolutional residual path), followed by a Leaky ReLU[54] activation (slope 0.1) and a 2×2 max pooling layer (stride 2) that halves spatial dimensions and doubles channel depth. Conversely, the decoder path consists of four upsampling blocks. Input to each upsampling block is a concatenation of the previous upsampled output and corresponding feature maps from the encoder, which are first processed by an attention gate (three convolutional layers and a sigmoid operation) to highlight salient features. These upsampling blocks then perform 2× bilinear resizing and utilize a three-convolutional-layer residual block to reduce channel count by a factor of four. The network concludes with the final upsampling block, followed by another three- convolutional-layer residual block and a single convolutional layer, ultimately reducing the channels to 3 to match the ground truth image format.

The discriminator network, as shown in Figure 7(c), is designed to distinguish between the virtually stained images and the actual ground truth images. It begins by processing an input image through a single convolutional layer to create a 64-channel tensor, immediately followed by a Leaky ReLU activation function. This tensor then passes through five successive residual blocks, each composed of two convolutional layers. Within each block, the second convolutional layer uses a stride of 2, which achieves 2× downsampling of the spatial dimensions while simultaneously doubling the number of feature channels. After these feature extraction blocks, a global pooling layer aggregates the learned features, which are then fed into two dense (fully-connected) layers to ultimately produce a probability score indicating whether the input image is likely a real ground truth image or not.

The registration module[55] is built upon a U-Net architecture, similar to the generator but with key differences. It features seven pairs of downsampling and corresponding upsampling blocks, with each block in both paths being integrated with a residual block to facilitate feature propagation. At the deepest point of the network, following the final downsampling stage, a convolutional layer doubles the feature channels. This is succeeded by a series of three consecutive residual blocks for further feature refinement. Subsequently, another convolutional layer halves the channel count, preparing the features for the upsampling sequence. The network culminates in an output layer that utilizes a single convolutional layer to condense the feature channels to two, directly corresponding to the two normal components of the displacement matrix it aims to predict. This registration module, together with the discriminator, was only used during the training phase to learn more accurate virtual staining models.

Image patches with $512 \times 512$ pixels were used during the training of the models, which were obtained by randomly cropping the training images. Also, several data augmentation methods, including random image rotations (0, 90, 180, and 270 degrees) and random flipping, were applied to these image patches. Adam optimizer[56] was used to optimize the generator, discriminator, and registration modules (with the learning rates of $2 \times 10^{-5}$, $2 \times 10^{-6}$ and $2 \times 10^{-6}$, respectively). The batch size was set to 4. The generator parameters were updated four times for each update of the discriminator parameters. The training duration of the virtual staining models was approximately 72 hours. A computer with GeForce RTX 3090 Ti GPUs, 256GB of random-access memory (RAM), and an Intel Core i9 central processing unit (CPU) was used during training and blind evaluation of the models. These deep neural network models were implemented using Python version 3.12.0 and PyTorch[57] version 1.9.0 with CUDA toolkit version 11.8.

## Supplementary Information

- Neural network architecture and training loss function details

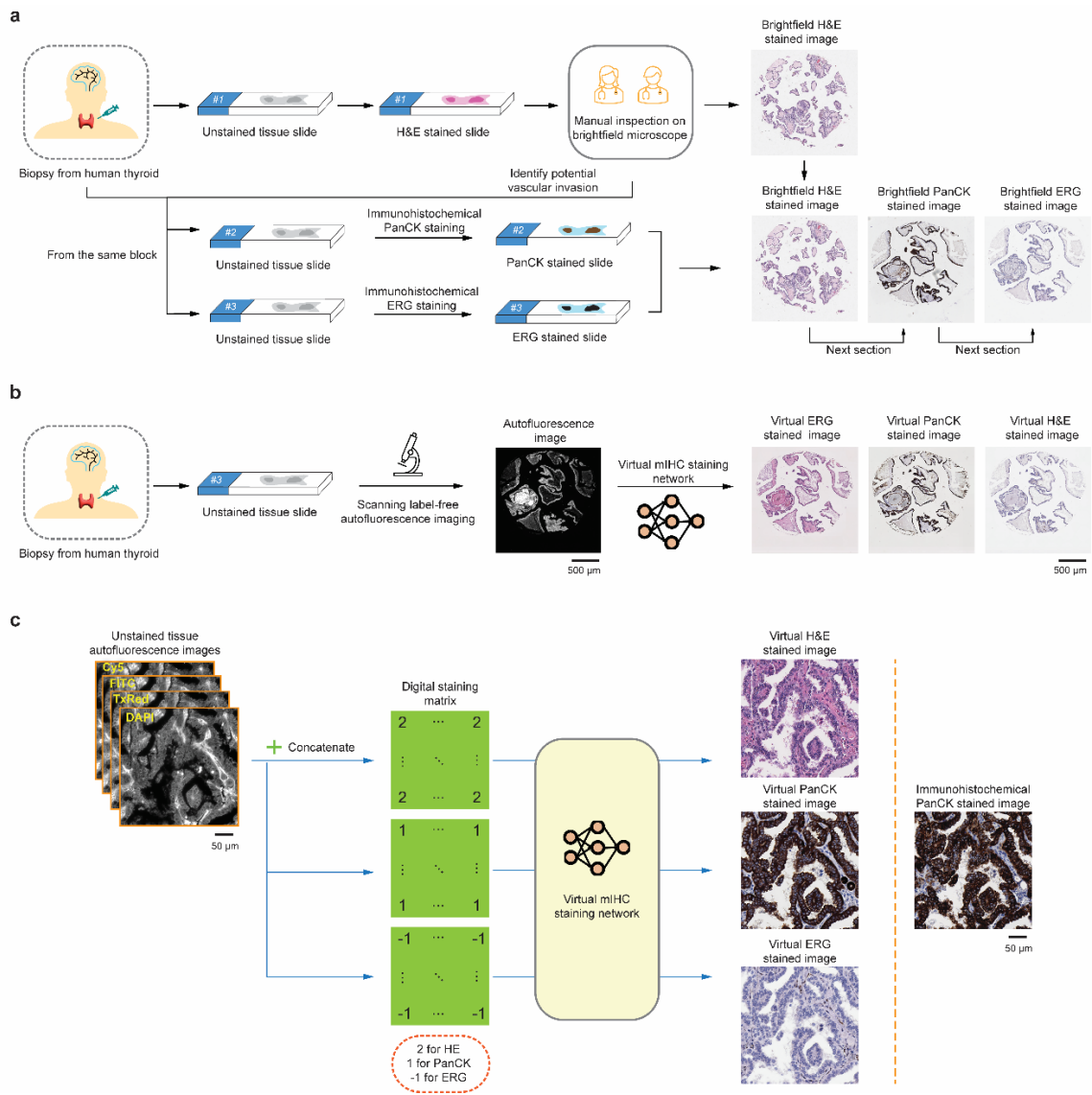

**Figure 1**. **Virtual multiplexed IHC (mIHC) staining of label-free thyroid tissue for vascular invasion assessment**. (a) Clinical workflow for assessing vascular invasion in human thyroid tissue. (b) Workflow of virtual mIHC staining using label-free autofluorescence imaging of unstained thyroid tissue. (c) Virtual staining network using autofluorescence inputs and a digital staining matrix to generate brightfield-equivalent images for H&E, PanCK, and ERG stains.

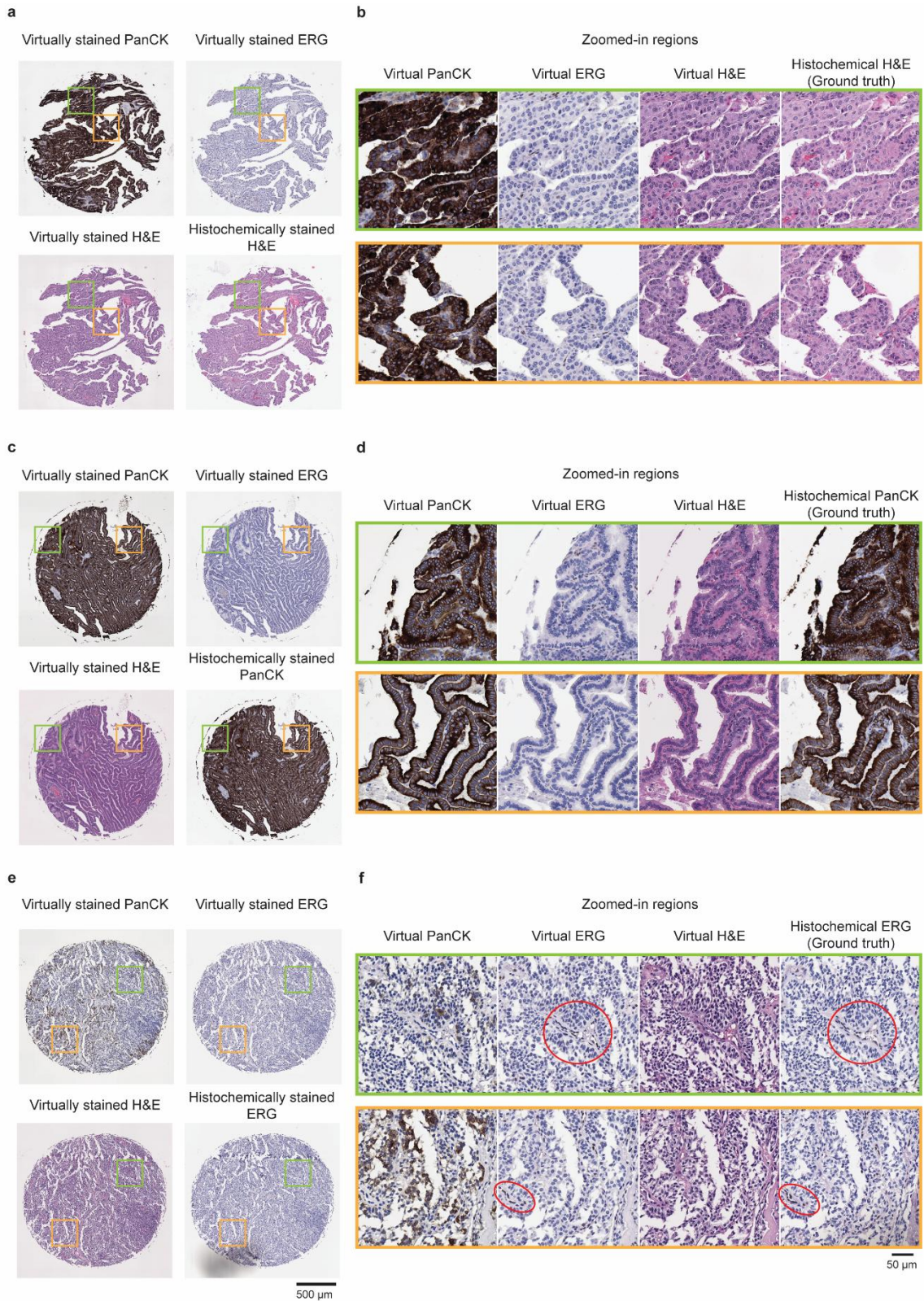

**Figure 2. Visual comparison between the virtual multiplexed stains of label-free thyroid sample and their histochemically stained counterparts.** (a, c, e) Virtually stained H&E, PanCK, and ERG images of

three representative label-free thyroid TMA cores. For each core, the corresponding histochemically or immunohistochemically stained reference image (H&E, PanCK, or ERG) is also shown. (b, d, f) Zoomed-in views of the selected regions (green and orange boxes) in (a, c, e), highlighting detailed comparisons between virtual and histochemical staining for each stain type. The H&E virtual slides closely reproduce the tissue morphology and color contrast seen in the histochemical reference, enabling the evaluation of cellular size and shape, nuclear chromatin texture and stromal organization. Virtual IHC staining for PanCK demonstrates cytoplasmic staining of epithelial cells, allowing the evaluation of epithelial distribution and integrity. ERG IHC staining shows a typical nuclear pattern, as highlighted in red circled regions in (f), enabling the assessment of vascular and endothelial structures.

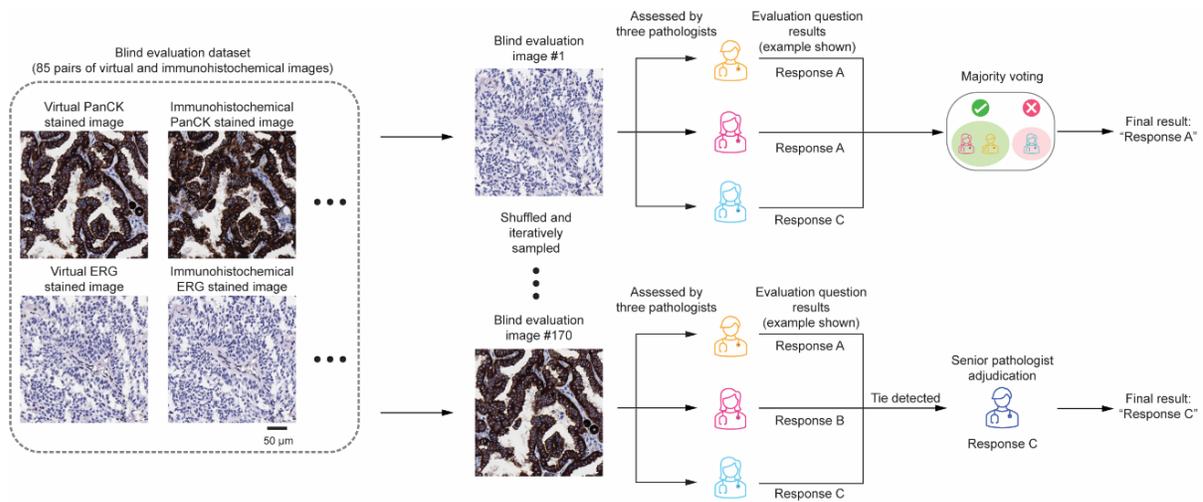

**Figure 3. Overview of the study design incorporating board-certified pathologists' review.** A total of 85 pairs of virtual and immunohistochemically stained images (170 images in total) were randomly shuffled and iteratively presented to three pathologists in a blinded fashion. Each image was independently assessed, and responses to predefined evaluation questions were recorded. The final evaluation result was determined by majority voting. In cases where a tie occurred, a 4th senior pathologist was consulted to adjudicate and provide the final decision.

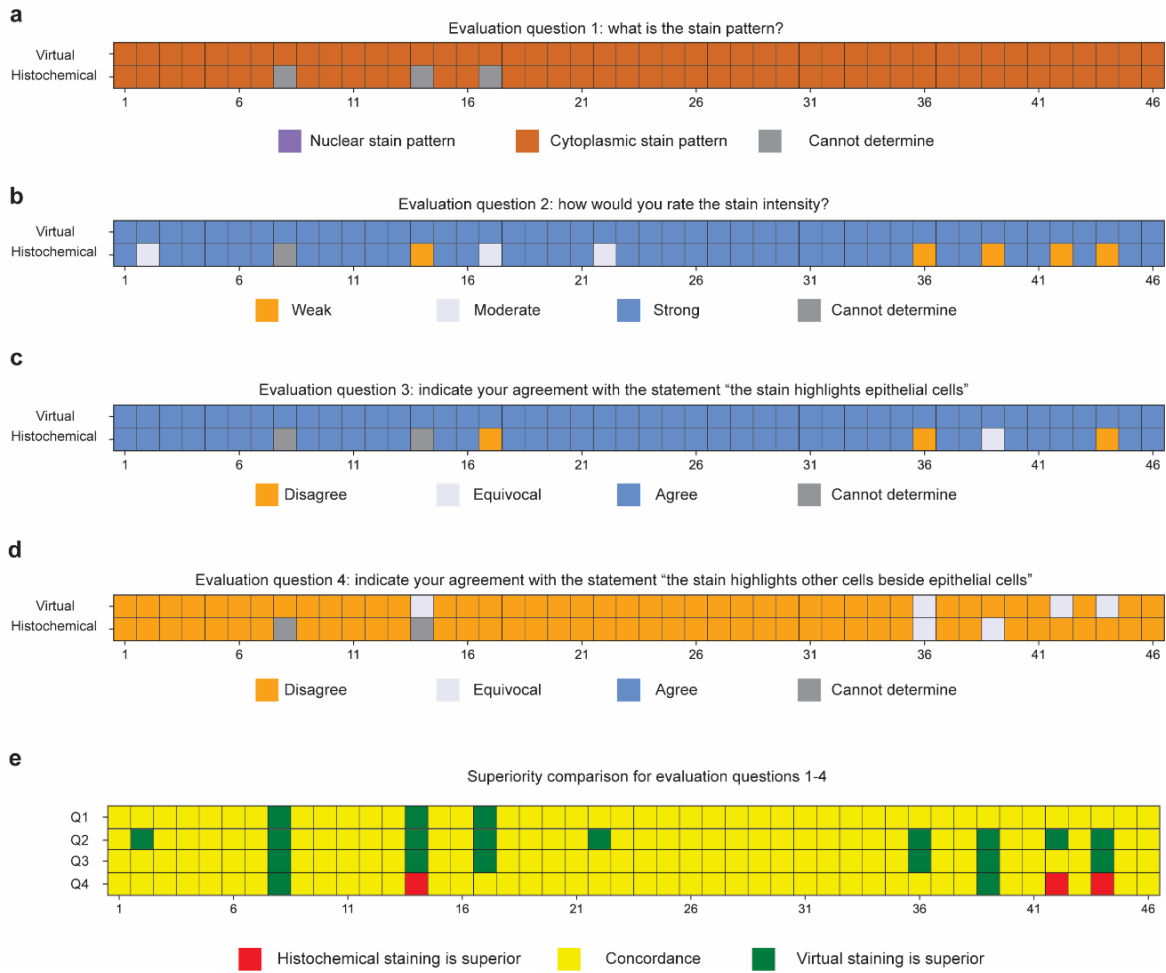

**Figure 4. Visualization of pathological evaluation results for PanCK stain.** (a–d) Evaluation results for questions 1–4 (as defined in Figure 3) for each virtually and histochemically stained PanCK image pair. The questions assessed (a) stain pattern, (b) stain intensity, (c) agreement on whether the stain highlights epithelial cells, and (d) whether the stain highlights non-epithelial cells. (e) Superiority comparison based on evaluation questions 1–4, indicating whether the virtually stained or histochemically stained PanCK image was rated superior, or if the responses were concordant.

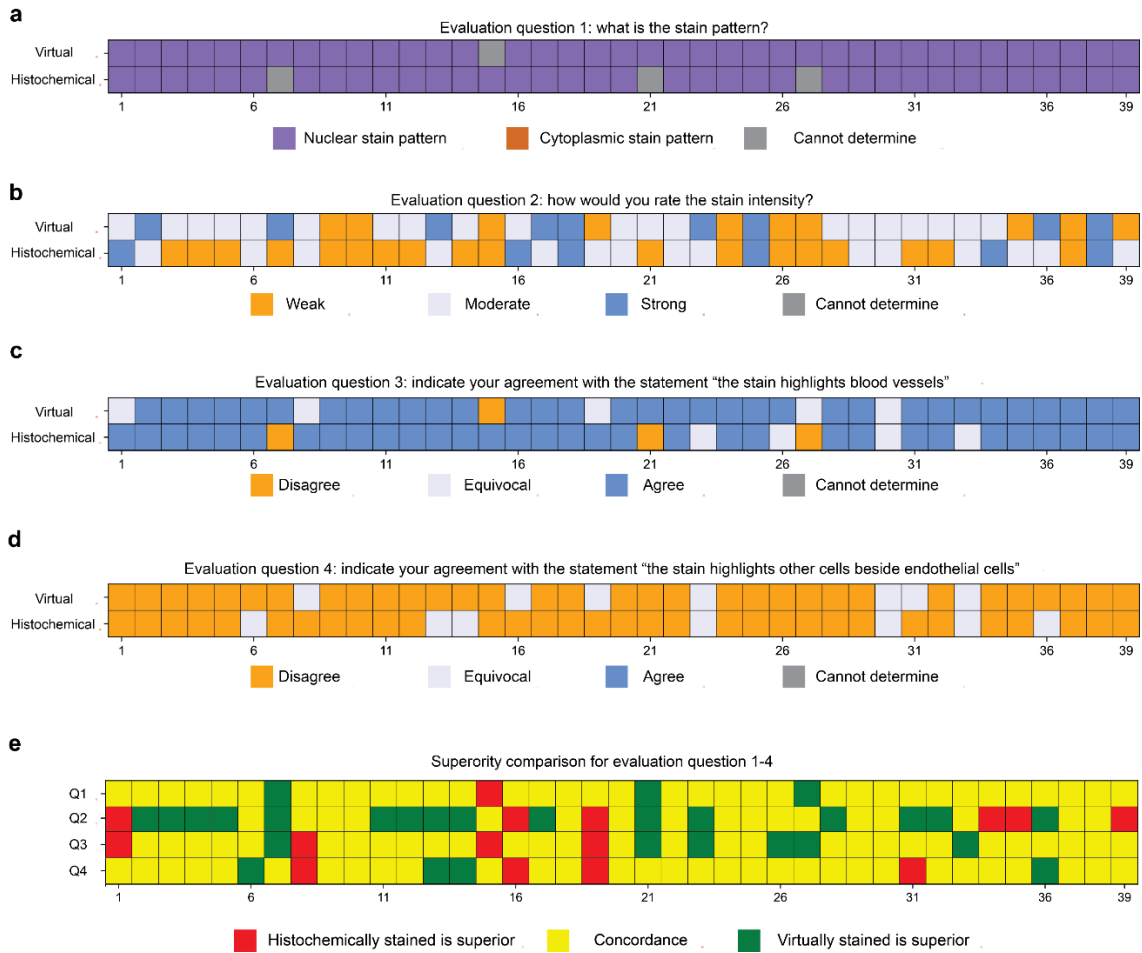

**Figure 5. Visualization of pathological evaluation results for ERG stain.** (a–d) Evaluation outcomes for questions 1–4 (as defined in Figure 3) for each virtual and histochemically stained ERG image pair, shown per case. Responses include (a) stain pattern, (b) stain intensity, (c) agreement on endothelial cell highlighting, and (d) agreement on staining of non-endothelial cells. (e) Superiority comparison for each evaluation question across all cases, indicating whether the virtually stained or histochemically stained image was rated superior, or if the results were concordant.

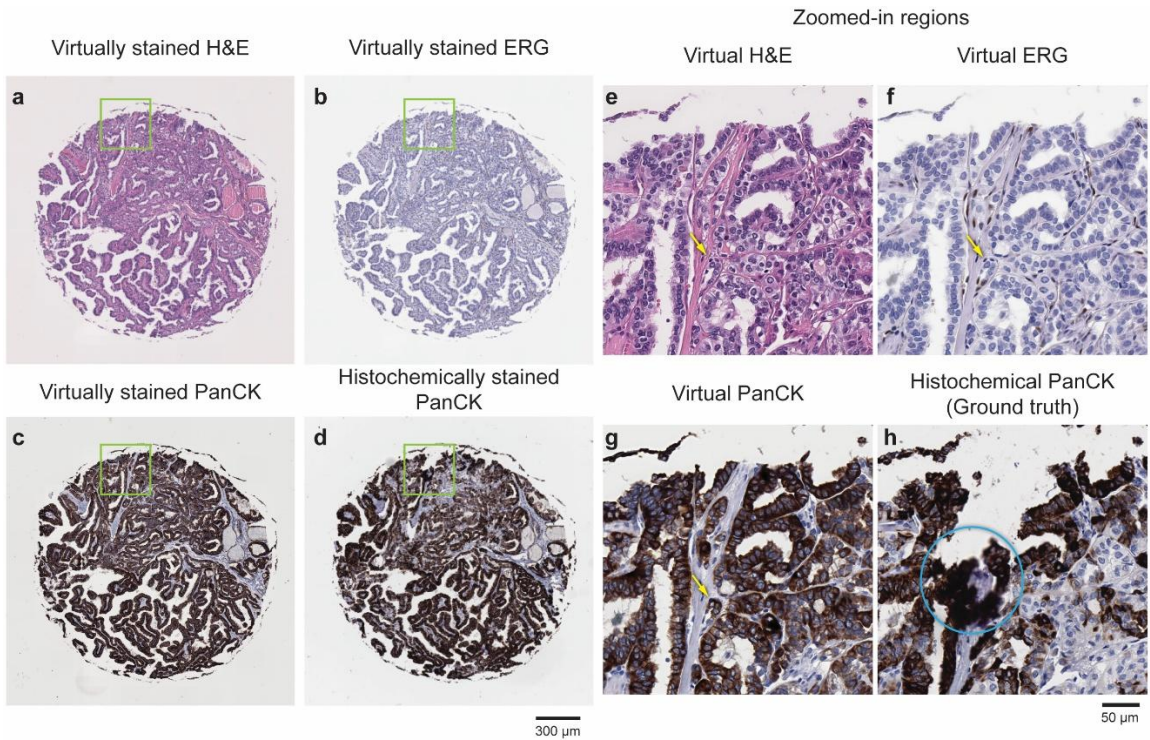

**Figure 6. Example of virtual mIHC staining used to identify vascular invasion**. (a–c) Virtually stained H&E, ERG, and PanCK images of a label-free thyroid TMA core from a patient diagnosed with metastatic disease. (d) Corresponding histochemically stained PanCK image of the same TMA core. (e–h) Zoomed-in views of the regions outlined in (a–d). As identified by a board-certified pathologist (N.P.), the yellow arrows indicate a region of vascular invasion, and the blue circled region indicates an area with non-specific staining, as part of the histochemical PanCK.

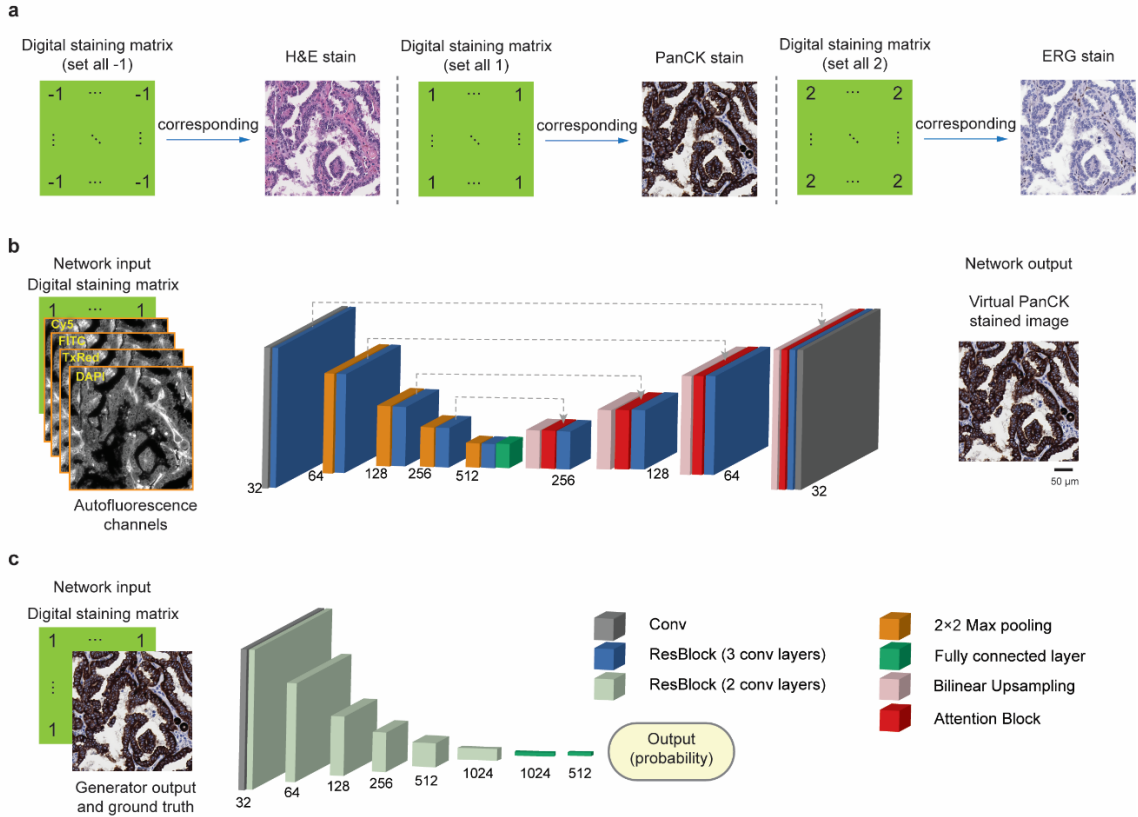

**Figure 7. Network architecture of the virtual mIHC staining model**. (a) Digital staining matrix configurations used during both the training and testing phases to generate multiplexed virtual stains: all elements set to –1 for H&E, 1 for PanCK, and 2 for ERG. (b) Detailed architecture and building blocks of the generator. (c) Detailed architecture and building blocks of the discriminator.